%% file: keps.tex
\documentclass[twocolumn,twoside,preprintnumbers,amsmath,amssymb,aps,nobibnotes,nofootinbib]{revtex4} 
\pdfoutput=1

\usepackage{graphicx}
\graphicspath{{Figs/}}
\usepackage{dcolumn}
\usepackage{bm}
\usepackage[a4paper, top=1.0in, bottom=1.0in, left=1.0in, right=1.0in]{geometry}
\usepackage{multirow}

\newcommand{\<}{\langle}
\renewcommand{\>}{\rangle}
\newcommand{\beq}{\beqn}
\newcommand{\eeq}{\eeqn}
\newcommand{\beqn}{\begin{eqnarray}}
\newcommand{\eeqn}{\end{eqnarray}}
\def\sla#1{\setbox0=\hbox{$#1$}\dimen0=\wd0
      \setbox1=\hbox{/} \dimen1=\wd1 \ifdim\dimen0>\dimen1
      \rlap{\hbox to \dimen0{\hfil/\hfil}} #1                        \else
      \rlap{\hbox to \dimen1{\hfil$#1$\hfil}}
      /   \fi}

\newcommand{\nn}{\nonumber}
\newcommand{\ov}{\overline}

\newcommand{\D}{\Delta}
\newcommand{\eps}{\epsilon}
\newcommand{\epsp}{\epsilon^\prime}
\newcommand{\epe}{\epsilon^\prime/\epsilon}
\newcommand{\Deps}{\Delta_\eps}
\newcommand{\la}{\lambda}

\newcommand{\bk}{\hat B_K}
\newcommand{\keps}{\kappa_{\epsilon}}
\newcommand{\ovkeps}{{\overline\kappa}_{\epsilon}}

\newcommand{\mc}{\mathcal}
\newcommand{\re}{{\rm Re}}
\newcommand{\im}{{\rm Im}}
\newcommand{\Lms}{\Lambda^{(4)}_{\overline{\rm MS}}}
\newcommand{\BS}{B_6^{(1/2)}}
\newcommand{\BE}{B_8^{(3/2)}}
\newcommand{\mev}{{\rm MeV}}
\newcommand{\cw}{\cos \theta_w}

\newcommand{\phiNP}{\phi_{\rm NP}}

\long\def\symbolfootnote[#1]#2{\begingroup%
 \def\thefootnote{\fnsymbol{footnote}}\footnote[#1]{#2}\endgroup}

\newcommand{\noi}{\noindent}

\begin{document}

\preprint{TUM-HEP-708/09}
\preprint{CERN-PH-TH/2009-013}
\title{\boldmath On the consistency between the observed amount of CP violation\\in the $K$- and $B_d$-systems within minimal flavor violation}

\author{Andrzej~J.~Buras$^{a,b}$}
\author{Diego~Guadagnoli$^{a,c}$}
\affiliation{%
\vspace{0.3cm}
$^a$Physik-Department, Technische Universit\"at M\"unchen,
D-85748 Garching, Germany\vspace{0.1cm}\\
$^b$TUM Institute for Advanced Study, Technische Universit\"at M\"unchen,
D-80333 M\"unchen, Germany\vspace{0.1cm}\\
$^c$CERN, Theory Division, CH-1211 Geneva 23, Switzerland
}%

\date{\today}

\begin{abstract}

\noi 
We reappraise the question whether the Standard Model, and Minimal Flavor Violating (MFV) models at large, can simultaneously describe the observed CP violation in 
the $K$- and $B_d$-systems.
We find that CP violation in the $B_d$-system, measured most precisely through $(\sin 2 \beta)_{J/\psi K_s}$, implies $|\eps_K^{\rm SM}| = 1.78(25) \times 10^{-3}$ 
for the parameter $\eps_K$, measuring indirect CP violation in the $K$-system, to be compared with the experimental value $|\eps_K^{\rm exp}| = 2.23(1) \times 10^{-3}$. 
To bring this prediction to $1\sigma$ agreement with experiment, we explore then the simplest new-physics possibility not involving new phases, namely that of MFV 
scenarios with no new effective operators besides the Standard Model ones. We emphasize the crucial input and/or measurements to be improved in 
order to probe this case. In particular we point out that this tension could be removed in this framework, with interesting signatures, e.g. correlated 
suppression patterns for rare K decay branching ratios. On the other hand, MFV contributions from new operators appear, in the calculable case of the MSSM, to worsen 
the situation. We finally explore some well-motivated new-physics scenarios beyond MFV, like those involving generic new contributions in $Z$-penguins.
\end{abstract}


\maketitle

\section{\boldmath Introduction}

Forty-four years after its discovery in the decay $K_L \to \pi \pi$ \cite{CCFT64}, CP violation leaves plenty of open questions. 
In the Standard Model (SM) CP violation is generated by the physical phase appearing in the CKM matrix, that in turn governs all flavor-violating 
interactions. While this picture of flavor and CP violation cannot be viewed as a fundamental theory of flavor, it turns out to be a very successful 
parameterization of intergenerational quark interactions, in which the hierarchies in CP-violating phenomena predicted in $K$, $B_d$, $B_s$ and $D$ decays
are strongly correlated with the hierarchies of CP-conserving, but flavor-violating decays \cite{CERNJan08}. At the root of these correlations is the uniqueness of 
the CKM phase. In extensions of the SM, because of the natural presence of new flavor-violating interactions as well as CP-violating phases, such a delicate pattern 
is in general badly destroyed. Therefore probing it to the best possible accuracy provides one of the most crucial SM tests.

Our knowledge of CP-violating phenomena is based on the following measurements: 
{\em (i.)} the parameter $\eps_K$ (indirect CP violation) in $K_L \to \pi \pi$ and $K_L \to \pi \ell \nu$ decays; 
{\em (ii.)} the parameter $\epsp$ (direct CP violation) in $K_L \to \pi \pi$ decays; 
{\em (iii.)} the parameter $\sin 2 \beta$ (CP violation in the interference between mixing and decay), very precisely determined from $B \to J/\psi K_S$ 
decays and with still significant theoretical and experimental uncertainties in several additional modes; 
{\em (iv.)} direct CP violation in various hadronic $B$ decays, again with still substantial uncertainties \cite{KN-PDG}.

On the other hand, no evidence exists to date for CP violation in the $D$ and $B_s$ systems, which in the SM is predicted to be tiny, so that precisely
these two systems would offer the most crucial probes of non-SM CP-violating effects. 

Due to the theoretical and/or experimental uncertainties involved, it may 
still take some time until measurements in {\em (ii.)} and {\em (iv.)} above become important as tests of the CKM picture at the quantitative level. 
Instead, a major insight on the CKM correlation between $\eps_K$ and $\sin 2 \beta$ could become possible in the coming years through
\begin{itemize}
\item[1.] an improved determination of $\sin 2 \beta$ and in particular of the CKM angle $\gamma$ through tree-level decays;
\item[2.] improved calculations of the non-perturbative parameter $\bk$, that crucially enters the formula for $\eps_K$.
\end{itemize}

Basing on existing analyses of the Unitarity Triangle (UT), the measured value of $\sin 2 \beta$, dominated by the measurement of the time-dependent asymmetry 
in $B \to J/\psi K_S$, and the value of $\eps_K$ are regarded as consistent with each other within the CKM picture of flavor and CP violation. It should however 
be stressed that this $\sin 2 \beta - \eps_K$ correlation is still far from being accurate at the theoretical level. Indeed, as seen in any plot of the UT, 
while the $\sin 2 \beta$ constraint in the $(\ov \rho - \ov \eta)$-plane is very strong, the corresponding one from $\eps_K$ is fairly weak. Confidence that 
the size of CP violation in the $B_d$-system ($\sin 2 \beta$) and in the $K$-system ($\eps_K$) are consistent with each other is only at the 15\% level. 
This fundamental test of consistency of CP violation across different generations is by the way the only one available at present.

In a recent paper \cite{BG} we have raised the possibility that the SM prediction of $|\eps_K|$ implied by the measured value of $\sin 2 \beta$ may be too 
small to agree with experiment. Two main ingredients, absent in the existing UT analyses to date, led to the above hypothesis:
\begin{itemize}
\item[{\em a.}] a decrease of $\bk$ to the value \cite{DJAntonio} (see also \cite{Allton})
\beq
\bk = 0.720 \pm 0.013 \pm 0.037~,
\label{BKRBC}
\eeq
lower by 5-10\% with respect to the values used in existing UT fits \cite{UTfit,CKMfitter};
\item[{\em b.}] the observation \cite{BG} that effects neglected in the usually adopted formula for $\eps_K$ amount to an additional suppression, 
that can be parameterized as a multiplicative factor, estimated within the SM as
\beq
\keps = 0.92 \pm 0.02~.
\label{keps}
\eeq
\end{itemize}
Because $\eps_K \propto \bk \keps$, the total suppression of $\eps_K$ with respect to the commonly adopted formulae is potentially of the order of 20\%.
These facts motivated us in \cite{BG} to look in more detail into the $\eps_K - \sin 2 \beta$ correlation, in particular at the $\bk$ range implied by the 
assumption that the correlation be fully described by the SM. It should be mentioned that our study has been inspired by a complementary analysis of Lunghi 
and Soni \cite{LunghiSoni-Bd}, who, assuming no NP in $\eps_K$ and using the value of $\bk$ from \cite{DJAntonio}, found even in the
limit $\keps=1$ values for $\sin 2\beta$ visibly larger than $(\sin 2 \beta)_{J/\psi K_s}$.

With present data, no statement above the $2\sigma$ level can be made \cite{BG,LunghiSoni-Bd,Lellouch-LAT08}. However, an improvement in the relevant input, 
e.g. an independent lattice determination of $\bk$ confirming point {\em (a)}, has in our opinion a concrete potential to uncover an inconsistency between 
$\eps_K$ and $\sin 2 \beta$ within the SM.

Purpose of the present paper is to provide additional arguments for the above possibility and to comment on how the $\eps_K - \sin 2\beta$ correlation 
--~along with additional observables in the flavor sector~-- is modified within the simplest extensions of the SM, within and beyond Minimal Flavor Violation 
(MFV).

\section{\boldmath $\eps_K$ in the Standard Model}

Let us first recall that within the SM
\begin{itemize}
\item[{\em i.}] $S_{J/\psi K_S} =\sin 2 \beta $ measures directly the phase $\beta$ (see \cite{CPS,Fleischer-s2b,GR} for corrections to this relation);
\item[{\em ii.}] with the implied precise value of $\beta$, $|\eps_K|$ can be predicted in terms of the remaining three parameters of the CKM matrix,
that we choose to be $|V_{us}|$, $|V_{cb}|$ and the UT side $R_t$, the rest of the parametric dependence being in the loop functions, in $\bk$ and in $\keps$.
\end{itemize}
\begin{table}[th]
\footnotesize
\center{
\begin{tabular}{|l|l|}
\hline
& \\
[-0.25cm]
$G_F = 1.16637 \cdot 10^{-5}$ GeV$^{-2}$            &     $\eta_{cc} = 1.43(23)$ \hfill \cite{HerrlichNierste} \\
$M_W = 80.398(25)$ GeV                              &     $\eta_{ct} = 0.47(4)$ \hfill \cite{HerrlichNierste} \\
$M_t = 172.6(1.4)$ 
GeV\symbolfootnote[1]{The ${\rm \ov{MS}}$ mass 
value $m_t(m_t) = 162.7(1.3)$ is derived 
using \cite{RunDec}.} \hfill \cite{CDF-D0-top}      &     $\eta_{tt} = 0.5765(65)$ \hfill \cite{BurasJaminWeisz} \\
$m_c(m_c) = 1.270(17)$ GeV                          &	  $F_K = 0.1561(8)$ GeV \hfill \cite{Flavianet}\\
$M_{B_d} = 5.2795(3)$ GeV                           &	  $M_{K^0} = 0.497614$ GeV \\
$M_{B_s} = 5.3663(6)$ GeV                           &	  $\D M_K = 0.5292(9) \cdot 10^{-2}/{\rm ps}$ \\
$\D M_d = 0.507(5)/{\rm ps}$                        &	  $|\eps_K| = 2.229(12) \cdot 10^{-3}$ \\
$\D M_s = 17.77(12)/{\rm ps}$ \hfill \cite{CDF-DMs} &	  $\keps = 0.92(2)$ \hfill \cite{BG}\\
$\xi_s = 1.21(4)$ \hfill \cite{Gamiz-LAT08}         &	  $\phi_\eps = 43.5(7)^\circ$ \\
$\lambda = 0.2255(7)$ \hfill \cite{Flavianet}       &	  $\epe = 1.65(26) \cdot 10^{-3}$ \hfill \cite{PDG08,Batley-epe,AlaviHarati-epe}\\
$|V_{cb}| = 41.2(1.1) \cdot 10^{-3}$                &	  $\sin 2 \beta = 0.675(26)$ \hfill \cite{HFAG08}\\
\hline
\end{tabular}
}
\caption{Input parameters. Quantities lacking a reference are taken from \cite{PDG08}.}
\label{tab:input}
\end{table}
From point {\em (ii)} and eq. (13) of \cite{BG} one easily gets
\beq
&&|\eps_K|^{\rm SM} = \keps C_\eps \bk |V_{cb}|^2 |V_{us}|^2 \times \nn \\
&&\hspace{1cm}\Bigl( \frac{1}{2} |V_{cb}|^2 R_t^2 \sin 2 \beta ~ \eta_{tt} S_0(x_t) \nn \\
&&\hspace{1cm}+ R_t \sin \beta~\left( \eta_{ct} S_0(x_c,x_t) - \eta_{cc} x_c \right) \Bigl)~, \nn \\
[0.2cm]
&&\mbox{with  } C_\eps = \frac{G_F^2 F_K^2 M_{K^0} M_W^2}{6 \sqrt 2 \pi^2 \D M_K}~,
\label{epsapprox}
\eeq
where the SM loop functions $S_0$ (see e.g. \cite{BurasLesHouches}) depend on $x_i = \ov m_i^2(m_i)/M_W^2$. The residual approximations involved in eq. (\ref{epsapprox}) are 
well below 1\%. Using the parametric input reported in table \ref{tab:input} (cf. \cite{BG}) --~implying $R_t = 0.914 \pm 0.031$ through $\Delta M_d/\Delta M_s$~-- 
and the result of \cite{DJAntonio} for $\bk$,\footnote{$\bk$ has been estimated by various other lattice collaborations \cite{BK-Aoki1997,BK-Aoki2004,BK-Aoki2005,
BK-Gamiz2006,BK-Aoki2008,BK-Nakamura2008}. We choose the value of \cite{DJAntonio} since the involved systematics should be minimal (cf. \cite{DJAntonio}, caption 
of fig. 4).} we find
\beq
|\eps_K|^{\rm SM} = (1.78 \pm 0.25) \times 10^{-3}~,
\label{epsKSM}
\eeq
to be compared with
\beq
|\eps_K|^{\rm exp} = (2.229 \pm 0.012) \times 10^{-3}~.
\label{epsKexp}
\eeq
The 15\% error in eq. (\ref{epsKSM}) can be understood most simply in terms of the three main sources of uncertainty in eq. (\ref{epsapprox}), namely
$\bk$, $|V_{cb}|^4$ and $R_t^2$, the latter two components entering the top-top contribution to $\eps_K^{\rm SM}$, that in turn constitutes about 75\% of the full 
result. A natural question is whether the discrepancy between eq. (\ref{epsKSM}) and eq. (\ref{epsKexp}) may be due to short-distance physics, which is encoded in 
the loop functions and in the $\keps$ factor. Correspondingly, in the next sections we will explore the kind of new physics required in $S_0$ and in $\keps$ to 
bring eq. (\ref{epsKSM}) to $1\sigma$ agreement with experiment, and the impact on other observables.
Needless to say, a simple solution to the tension between (\ref{epsKSM}) and (\ref{epsKexp}) is an increased value of $\sin 2\beta$, that would imply new phases 
in $B^0_d-\bar B^0_d$ mixing. This solution has been analyzed in detail already in \cite{BG,LunghiSoni-Bd} and we will not consider it here.

Barring all these possibilities, one is led to the conclusion that better agreement between (\ref{epsKSM}) and (\ref{epsKexp}) requires higher values for 
$\bk$, $R_t$ or $|V_{cb}|$. The fate of the test of the $\eps_K - \sin 2 \beta$ correlation within the SM depends crucially on these three inputs.

\section{\boldmath New physics in the $\Delta F = 2$ loop functions}\label{sec:loop}

Let us first address the possibility of a modification in the loop functions $S_0$, assuming that the mechanism of flavor violation (encoded in the CKM
matrix) along with the set of relevant operators stay the same as in the SM. This set of assumptions embodies what is called constrained Minimal Flavor Violation
(CMFV) \cite{Buras-UUT,Buras-Zakopane,BBGT}. Since the pure top contribution in $\eps_K^{\rm SM}$ (first term in the parenthesis of eq. (\ref{epsapprox})) 
amounts to roughly 75\% of the total, it is reasonable to assume that new-physics contributions affect mostly this part. Now, for eq. (\ref{epsKSM}) to
recover 1$\sigma$ agreement with eq. (\ref{epsKexp}), one needs under our assumptions a +10\% shift in $S_0(x_t)$. Would this shift be visible elsewhere? 
The function $S_0$ enters also the SM formulae for the mass differences in the $B_{d,s} - \ov B_{d,s}$ systems, respectively $\D M_{d,s}^{\rm SM}$. However, 
the latter still suffer from substantial uncertainties, exceeding 20\%, in the relevant lattice input $F^2_{B_q} \hat B_q$, $q=d,s$. As an example, taking 
$F_{B_s} \simeq 0.245$ GeV, $\hat B_s \simeq 1.30$ and $\xi_s \equiv (F_{B_s} \sqrt{\hat B_s})/(F_{B_d} \sqrt{\hat B_d}) \simeq 1.21$, and further including 
the assumed $\delta S_0 = +10\%$ shift in the $S_0$ function, parameterized as
\beq
S_0(x_t) \to S_0(x_t) (1 + \delta S_0)~,
\label{dS0}
\eeq
one would get the CMFV predictions
\beq
\label{DMdDMsCMFV}
&& \D M_d^{\rm CMFV} \approx (0.638 \pm 20\%)/{\rm ps}~,\nn \\
&& \D M_s^{\rm CMFV} \approx (21.6 \pm 20\%)/{\rm ps}~.
\eeq
Comparing with the experimental results reported in table \ref{tab:input}, one notices that both central values in eq. (\ref{DMdDMsCMFV}) exceed experiment by about 
20\%, but errors are also of this size. It is clear that sensitivity of $\D M_{d,s}^{\rm SM}$ to an $S_0$ shift will only be possible when the mentioned lattice 
input is controlled to a matching accuracy. In general, with lower errors on eqs. (\ref{DMdDMsCMFV}), increased values of $S_0$ would have to be compensated 
by decreased values of $F^2_{B_q} \hat B_q$ in order for the CMFV predictions to be in agreement with the experimental $\Delta M_{d,s}$ reported in table \ref{tab:input}.
\begin{figure}[tb]
\begin{center}
\includegraphics[width=0.48 \textwidth]{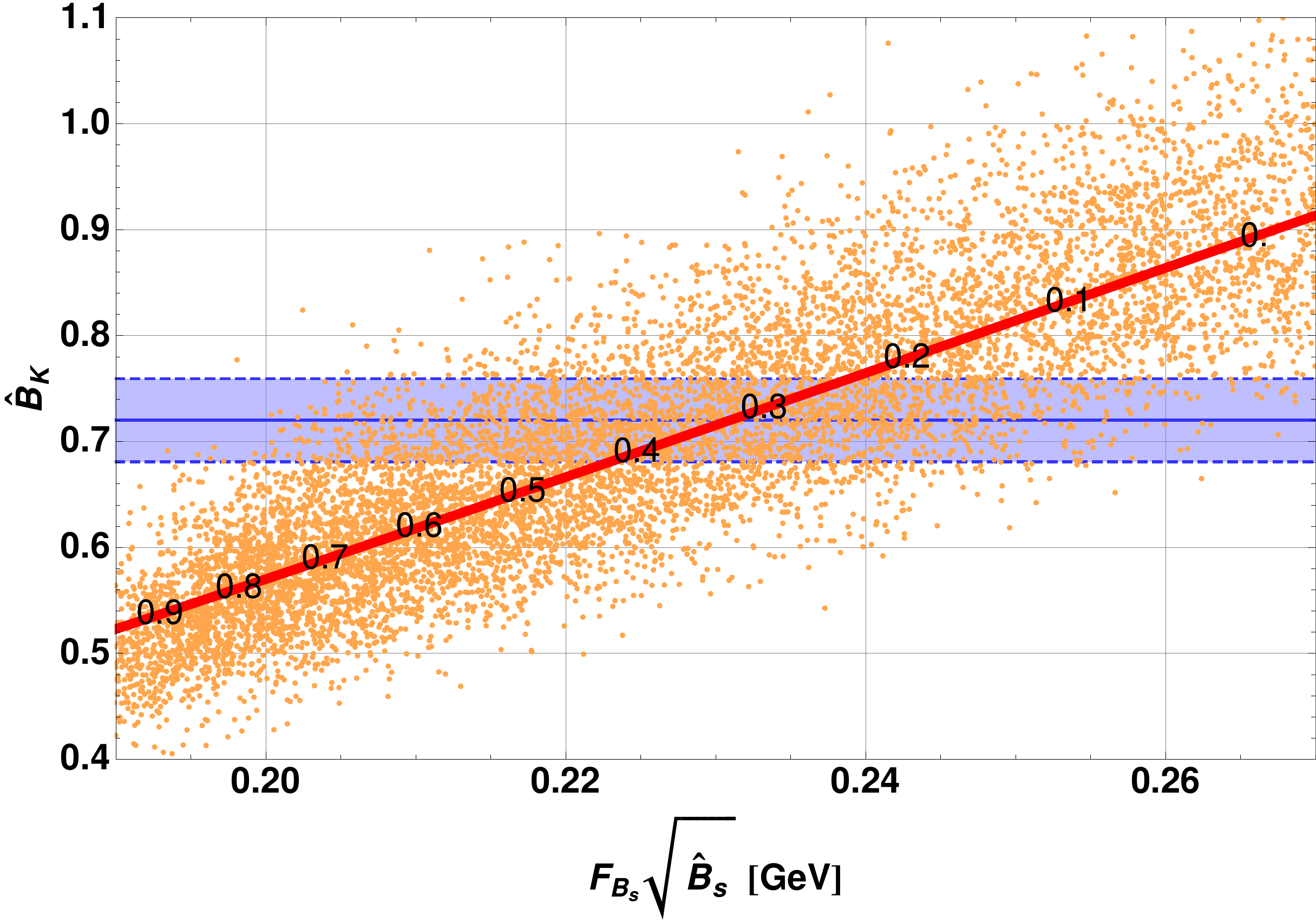}
\end{center}
\caption{\small\sl Values of $F_{B_s} \sqrt{\hat B_s}$ vs. $\bk$ required by $\D M_s$ and $\eps_K$ (see text for details). The scattered points are obtained by assuming
the theoretical input to $\D M_s$ and $\eps_K$ to be normally distributed around the values listed in table \ref{tab:input}. The red (solid) line corresponds to the case 
of central values on all input. Superimposed to the red line is the $\delta S_0$ shift. The horizontal band reports the $\bk$ range (\ref{BKRBC}).}
\label{fig:FsBvsBK}
\end{figure}
This point can be appreciated quantitatively in fig. \ref{fig:FsBvsBK}. This figure displays the values of $F_{B_s} \sqrt{\hat B_s}$ vs. $\bk$ required
by the agreement of the $\D M_s^{\rm CMFV}$ and $\eps_K^{\rm CMFV}$ predictions with the experimental data. The scattered points are obtained by assuming that the
theoretical input (other than $F_{B_s} \sqrt{\hat B_s}$ and $\bk$) obey normal distributions according to the values listed in table \ref{tab:input}. The case of 
central values on all the input is reported as a red (solid) line. Superimposed to the latter are also the values of $\delta S_0$ (see definition (\ref{dS0})).
The $\bk$ range (\ref{BKRBC}) is reported as well, as a horizontal band. For reference, unquenched determinations of $F_{B_s} \sqrt{\hat B_s}$ are in the ballpark of
0.245$-$0.281 GeV with about a 10\% quoted uncertainty \cite{Aoki03,Dalgic06,Albertus07} (see also refs. \cite{LT} and \cite{Gamiz-LAT08} for a collection of results).
As the simultaneous agreement with $\eps_K$ and $\Delta M_s$ corresponds to the overlap of the blue and red bands in fig. \ref{fig:FsBvsBK}, the downward shift of 
$F_{B_s}\sqrt{\hat B_s}$ mentioned before is clearly seen, although,~in view of the large lattice errors, it cannot be appreciated at present.

The {\em ratio} of $\D M_d^{\rm SM}$ to $\D M_s^{\rm SM}$ also affects the SM UT side $R_t$, with substantially smaller, O(5\%) lattice uncertainties, since
those on $\D M_d^{\rm SM}$ and $\D M_s^{\rm SM}$ are largely correlated. However, within CMFV, a shift in $S_0$ affects $\D M_d$ and $\D M_s$ universally
thereby exactly canceling in $R_t$ \cite{Buras-UUT}. Thus, in the case of CMFV models, the only route for prediction (\ref{epsKSM}) to get closer to 
(\ref{epsKexp}) via a shift in the loop functions can come from $\delta S_0>0$. Within CMFV models $\delta S_0>0$ is actually the most likely possibility 
\cite{BlankeBuras,ABG}.

More interesting can in principle be the case of a completely general MFV \cite{MFV}, where one just requires that any new flavor structure inherits from the SM Yukawa
couplings (see also \cite{chivukula-georgi,hall-randall}). In this framework, the occurrence of new contributions proportional to operators other than those relevant
within the SM is not forbidden, and indeed they arise in e.g. the two-Higgs-doublet extension of the SM \cite{MFV}, relevant also to the MSSM. With regards to meson
systems mass differences, the largest contributions from operators other than the SM (V$-$A)$\otimes$(V$-$A) structure are due to scalar operators. The latter, being 
proportional to the quark masses of the external states, are negligible in $\eps_K$ and affect differently $\D M_d$ and $\D M_s$, hence they are potentially visible 
in $R_t$. However, in the calculable case of the MSSM, MFV effects not accounted for by CMFV will shift $R_t$ {\em beneath} the SM value \cite{ABGW,FGH}, 
since their dominant impact is to add destructively to the SM contributions in $\D M_s$ \cite{BCRS01,IR} (for a very recent reappraisal of this issue, 
see \cite{Jager}). On the other hand, improved agreement with the $\eps_K$ constraint would require $R_t$ values {\em above} the SM one.

In short, the $\eps_K - \sin 2 \beta$ correlation can be improved with respect to the SM already by invoking MFV new-physics contributions universal to all meson
mixings, as in CMFV. This possibility is however tested at a level presently not better than 20\% and cries out progress in the $F^2_{B_q} \hat B_q$ estimations. 
If instead one is after MFV effects not accounted for within CMFV, i.e. from non-SM operators, then they would most likely come from SM extensions other
than the MSSM, as the latter appears to increase the tension between (\ref{epsKSM}) and (\ref{epsKexp}).

\section{\boldmath New physics in $\keps$}

Let us now address the possibility that $\keps$ be different from the value in (\ref{keps}), in particular higher, as required to recover $1\sigma$
agreement between eqs. (\ref{epsKSM}) and (\ref{epsKexp}). For the reader's convenience, we briefly summarize here the origin of this correction factor in $\eps_K$. 
The $\eps_K$ parameter can be calculated through the general formula \cite{BG}
\beq
\label{epsexact}
\eps_K = e^{i \phi_\eps} \sin \phi_\eps \left( \frac{\im (M^K_{12})}{\D M_K} + \xi \right)~,
\eeq
where
\beq
\xi = \frac{\im A_0}{\re A_0}~,
\eeqn
with $A_0$ the 0-isospin amplitude in $K \to \pi \pi$ decays, $M^K_{12} = \<K | \mc{H}^{\rm full}_{\D F = 2} | \ov K \>$, $\D M_K$ the 
$K - \ov K$ system mass difference, and the phase $\phi_\eps = (43.5 \pm 0.7)^\circ$ (see table \ref{tab:input}). The approximate $\eps_K$ formula typically 
used in phenomenological analyses can be recovered from (\ref{epsexact}) by setting $\phi_\eps = \pi/4$ and $\xi = 0$. Since deviations from $\phi_\eps = \pi/4$
and $\xi = 0$ can be regarded as perturbations, one can parameterize their combined effect as an overall factor $\keps$ in $\eps_K$, namely
\beq
\keps = \frac{\sin \phi_\eps}{1/\sqrt 2} \times \ovkeps~,
\label{kepsdef}
\eeq
with $\ovkeps$ parameterizing the effect of $\xi \neq 0$ through
\beq
\ovkeps = 1 + \frac{\xi}{\sqrt{2} |\eps_K|} \equiv 1+\Deps, 
\label{ovkepsdef}
\eeq
where $\Deps$ has been introduced for later convenience. As discussed in detail in \cite{BG}, a {\em direct} calculation of $\xi$ is subject at present to very large hadronic 
uncertainties, as no consensus exists on the value of the non-perturbative parameter $B_6$, describing QCD-penguin operators, that dominate $\xi$. Much more reliable is the 
indirect strategy where one evaluates the EW-penguin contribution to $\epe$ and uses the experimental $\epe$ value to determine $\xi$ \cite{FNAL-report-2002,AOV-BK,AOV-epsK}. 
Allowing for a 25\% error in this estimate, one arrives within the SM at $\keps = 0.92 \pm 0.02$ \cite{BG}, as given in eq. (\ref{keps}). 
Hence the like sign of the two corrections in eq. (\ref{kepsdef}) turns out to build up a $-$8\% total correction with respect to the approximate $\eps_K$ formula.

However, the EW-penguin contribution to $\epe$ can be affected by non-SM physics. Within the SM and for MFV models at large, the EW-penguin contributions are generally 
dominated by $Z$-penguin diagrams \cite{BCIRS}, so that the simplest expectation for new-physics contributions is a shift in the $Z$-penguin amplitude (see \cite{HW} for an 
updated discussion). We would like to address the question how this shift may alter $\xi$. This can be done with a strategy, to be described in the next paragraph, entirely 
analogous to the indirect route to $\xi$ mentioned above. In section \ref{sec:beyondMFV} we will comment on how this strategy deals with a more general modification 
in $\epe$ from new physics.

\input tabPBENDR.tex
We start from the following convenient formula for evaluating $\epe$ within the SM \cite{BL-PLB,Buras-Jamin}
\beq
\frac{\epsp}{\eps} = \im \lambda_t \cdot F_{\epsp}(x_t)~,
\label{epePBE}
\eeq
where $\lambda_t = V^*_{ts}V_{td}$, $x_t$ has been already introduced and $F_{\epsp}$ is given by
\beq
F_{\epsp}(x_t) &=& P_0 + P_X ~ X_0(x_t) + P_Y ~ Y_0(x_t) \nn \\
&+& P_Z ~ Z_0(x_t) + P_E ~ E_0(x_t)~,
\label{Fe}
\eeq
with $X_0$, $Y_0$, $Z_0$ and $E_0$ combinations of Inami-Lim functions \cite{Inami-Lim}. The coefficients $P_i$ in eq. (\ref{Fe}) are defined as \cite{BL-PLB,Buras-Jamin}
\beq
P_i = r_i^{(0)} + r_i^{(6)} R_6 + r_i^{(8)} R_8~.
\label{PiPBE}
\eeq
Here $r_i^{(0)}$, $r_i^{(6)}$ and $r_i^{(8)}$ encode the information on the Wilson-coefficient functions of the $\Delta S = 1$ effective 
Hamiltonian at the next-to-leading order \cite{BJLW-NLO,BJLW-ADM1,BJL-ADM2,CFMR-NLO}, and their numerical values for different choices of $\Lms$ at $\mu=m_c$ 
in the NDR renormalization scheme are displayed in table~\ref{tab:PBENDR}. On the other hand, $R_{6,8}$, defined as
\beq
&&R_6 \equiv \BS \left[ \frac{121~\mev}{m_s(m_c)+m_d(m_c)} \right]^2~,\nn \\
&&R_8 \equiv \BE \left[ \frac{121~\mev}{m_s(m_c)+m_d(m_c)} \right]^2~,
\label{R6R8}
\eeq
encode, through the `$B$-parameters' $\BS$ ($\BE$), the information on the operator matrix elements $\< Q_6 \>_0$ ($\< Q_8 \>_2$) between a $K$-meson and
a $\pi \pi$-state with isospin $I$=0 ($I$=2). Eqs. (\ref{epePBE})-(\ref{PiPBE}) assume the $\Delta S = 1$ operator basis $Q_{1-10}$ (see \cite{BJL}) wherein 
$Q_6$ ($Q_8$) represents the most important QCD-penguin (EW-penguin) operator. On the impact of the additional magnetic penguins $Q_{11,12}$ we will add 
comments in the next section. Concerning $R_8$, we assume the reasonable range 
\beq
R_8 = 1.0 \pm 0.2~,
\label{R8val}
\eeq
that encompasses various estimates reviewed in \cite{Buras-Jamin}.
On the other hand, in view of the mentioned huge theoretical uncertainties, we make no assumption on $R_6$. Its range, necessary for the estimation of 
$\xi$, hence $\keps$, will instead be extracted indirectly by demanding equality of the theoretical $\epe$ formula with $(\epe)_{\rm exp}=(1.65\pm 0.26)
\times 10^{-3}$ (see table \ref{tab:input}), within its $1\sigma$ range.

More explicitly, once the $R_6$ range has been estimated, the entailed range for the correction $\Deps$, hence $\xi$ (see eq. (\ref{ovkepsdef})), can
be obtained from the following approximate, but quite accurate formula
\beq
\Deps \approx - \frac{1}{\omega} \im \lambda_t \cdot F_{\epsp}(x_t)|_{R_8 \to 0}
\label{DepsNUM}
\eeq
where $\omega = \re A_2 / \re A_0 = 0.045$. 
In order to derive this approximate expression for $\Deps$ let us recall the basic formula for $\epe$ (see e.g. \cite{BurasLesHouches})
\beq
\label{epe_basic}
\frac{\epsp}{\eps} = - \omega \Deps (1 - \Omega)~,
\eeq
where $(- \omega \Deps)$ represents by definition the sum of the $\D I = 1/2$ contributions to $\epe$, whereas $\Omega$ is the absolute value of the ratio between the 
$\D I = 3/2$ and the $\D I = 1/2$ contributions. We note that the r.h.s. of eq. (\ref{DepsNUM}) includes in the $\Deps$ estimate the contributions from the coefficients 
$r_i^{(0)}$ (see eq. (\ref{PiPBE})), that consist of a $\D I = 3/2$ component along with the $\D I=1/2$ one. The former component is not separated away in eq. (\ref{DepsNUM}). 
Using the results of ref. \cite{BJL}, one can however convince oneself that this approximation amounts to overestimating $|\Deps|$ by less than 10\%, even for O(50\%) 
new physics in $Z$-penguins (i.e. $\delta C = 0.5$, see below). Therefore, effectively, the limit $R_8 \to 0$ in the $P_i$ coefficients (\ref{PiPBE}) corresponds to 
$\Omega \to 0$ in (\ref{epe_basic}), hence the possibility to estimate $\Deps$ from the simple relation (\ref{DepsNUM}).

With this strategy at hand, we can now study how $\xi$ may be affected by new physics in $Z$-penguin contributions. The latter arise from the $\ov s Z d$ effective Lagrangian 
interaction, that reads ('t Hooft-Feynman gauge)
\beq
\label{LZ}
\mc{L}^{Z} = \frac{G_F}{\sqrt 2} \frac{g_2}{2 \pi^2} \frac{M_W^2}{\cw} ~Z_{ds}~ \ov s (\gamma_\mu)_L d ~Z^\mu ~+ {\rm h.c.}~,~~
\eeq
with the complex `coupling' $Z_{ds}$ given in the SM by
\beq
Z_{ds}^{\rm SM} = \la_t C_0(x_t)~.
\label{ZdsSM}
\eeq
One can now parameterize the presence of non-SM contributions in $Z_{ds}$ through the replacement \cite{BS,BFRS}
\beq
Z_{ds}^{\rm SM} \rightarrow Z_{ds} = \la_t C_0(x_t)(1 + \delta C e^{i \phiNP})~,
\label{ZdsNP}
\eeq
with arbitrary $\delta C$ and $\phiNP$. It should be remarked that, since the interaction in eq. (\ref{LZ}) is gauge-dependent, so is the coupling $Z_{ds}$ in 
eq. (\ref{ZdsNP}). In the SM, this gauge-dependence is rather weak, as it enters only in terms that are subleading in $m_t$ and is canceled in the functions 
$X_0, Y_0$ and $Z_0$ (eq. (\ref{Fe})), which are linear combinations of the gauge-dependent $C_0$ and other photon-penguin and box diagrams \cite{BBH}. Since in any
known extension of the SM the latter diagrams receive subdominant contributions with respect to those affecting $Z$-penguins, we expect that the gauge-dependence 
of new-physics contributions to $Z_{ds}$ be also very weak and that it be a very good approximation to parameterize the new-physics contributions by the modification 
of $Z_{ds}$ only \cite{BS}. Arguments for new physics modifying dominantly $Z$-penguins are given in \cite{BCIRS}.

\begin{figure*}[tb]
\begin{center}
\includegraphics[width=0.48 \textwidth]{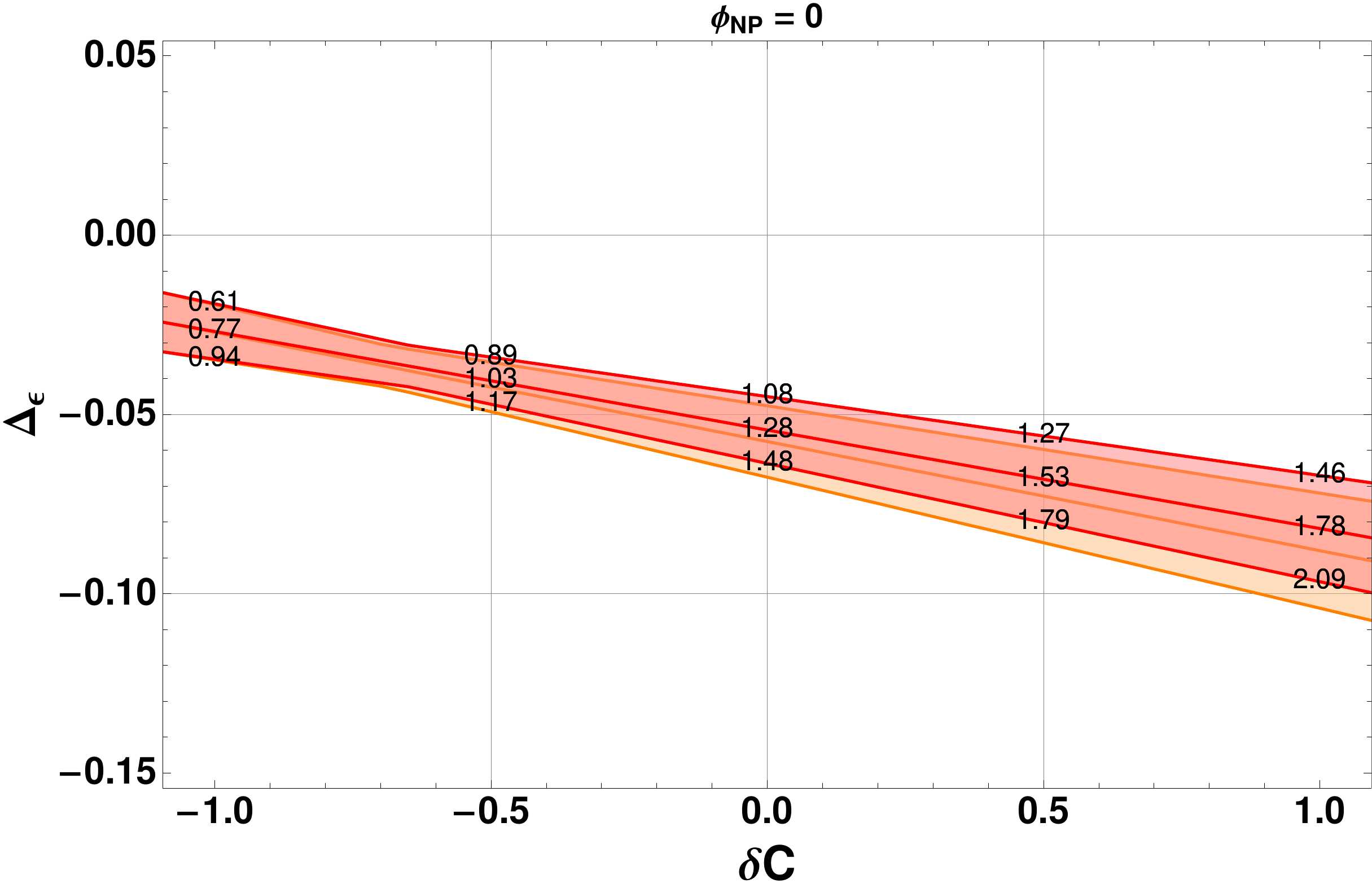} \phantom{xx} \includegraphics[width=0.48 \textwidth]{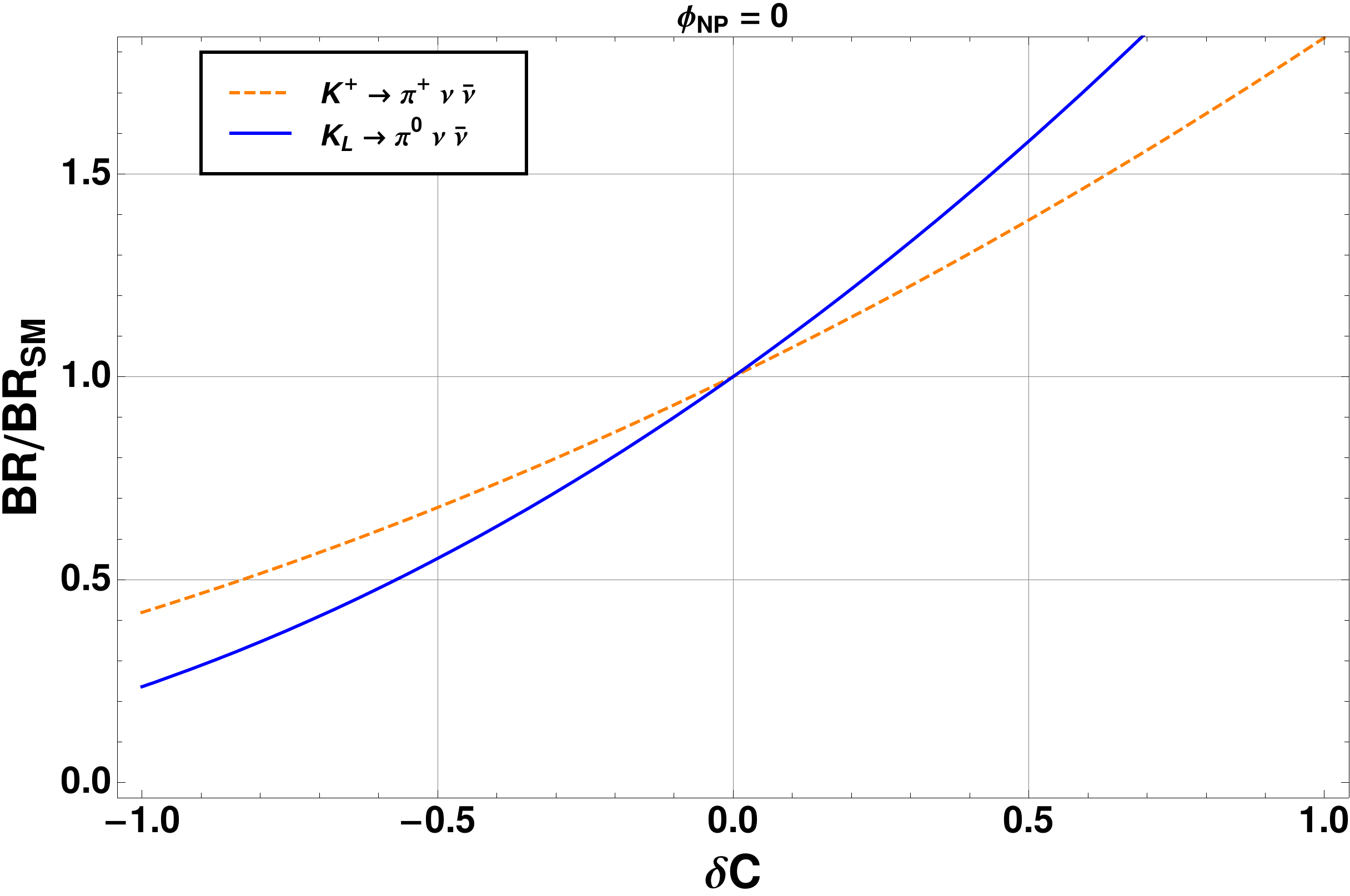}
\end{center}
\caption{\small\sl Left panel: Correction $\Delta_\eps$ as a function of the new-physics shift $\delta C$ in the $\ov s Z d$-vertex and $\phiNP = 0$. The darker (red) 
area refers to the choice $\Lms = 310$ MeV for the coefficients in tab. \ref{tab:PBENDR}. The lighter (orange) area corresponds instead to the choice $\Lms = 370$ MeV.
Displayed on the $\{0,\pm 1\}\sigma$ solid lines are also the values of $R_6$ corresponding to the given $\Deps$.
Right panel: Rate of enhancement in the branching ratios for the $K \to \pi \nu \ov \nu$ decays as a function of $\delta C$.}
\label{fig:CvsD}
\end{figure*}
To study the impact of the new-physics modification (\ref{ZdsNP}) on $\epe$, and in turn on $\xi$, let us first focus on the case of CMFV, where one additionally demands 
$\phi_{\rm NP} = 0$. The left panel of fig. \ref{fig:CvsD} shows the modification of a shift $\delta C \in [-1, 1]$ on $\Deps$, as defined in eq. (\ref{ovkepsdef}). 
For $\delta C = 0$, one can read $\Deps \approx -0.06$ \cite{BG}. Note that the chosen range for $\delta C$ is quite generous, taking into account the constraints
implied within MFV by other flavor observables \cite{bobeth} as well as by $Z \to b \ov b$ pseudo-observables \cite{HW}. In particular, positive shifts in $\delta C$, 
suppressing $\keps$ even further below unity are of no interest in this discussion.

We observe that, as expected, in order to increase $\keps$, or equivalently $\xi$, while keeping the experimental value of $\epe$ fixed, the magnitude of EW contributions 
to $\epe$ has to be decreased with respect to the SM case. This is apparent by noting, from table \ref{tab:PBENDR}, that the main contributions 
to QCD penguins (dominating $\xi$) and EW penguins, respectively $r^{(6)}_0$ and $r^{(8)}_Z$, come with opposite signs. From the left panel of the figure, one can note that, 
for $\keps$ to be outside the range in eq. (\ref{keps}), new physics in EW-penguins must be non negligible with respect to the SM contribution. For example, even a $\delta C$ 
shift as large as $-0.5$ would imply $\ovkeps \simeq 0.96$, whence, using eq. (\ref{kepsdef}), one would arrive at $\keps \simeq 0.93$.

We observe in addition that the new physics required to increase $\keps$ would generally suppress the branching ratios for rare $K$ decays. With our parameterization 
(\ref{ZdsNP}), this can be explored numerically by using the formulae of ref. \cite{BRS} (with parametric input taken from \cite{BGHN1,BGHN2,IMS,IMPRS}). 
The rate of suppression is displayed in the right panel of fig. \ref{fig:CvsD} for 
the decays $K^+ \to \pi^+ \nu \ov \nu$ and $K_L \to \pi^0 \nu \ov \nu$. A suppression on $K^+ \to \pi^+ \nu \ov \nu$ seems disfavored in the light of present 
knowledge \cite{Kp-exp} but data are definitely premature to draw any conclusion on this point.

As a further remark, even in the case where EW-penguin contributions are suppressed to zero, one would have $\keps = 0.94 \pm 0.01$. The decrease in the error in this case
is related to the fact that, in the absence of EW contributions to $\epe$, the relative error on $\xi$ is the same as that in $(\epe)_{\rm exp}$.\footnote{Figure \ref{fig:CvsD}
shows that the point with minimum $\Deps$ error is at $\delta C \approx -0.7$: this is where the EW-penguin contributions are exactly zero, thereby eliminating the $R_8$ 
contribution to the $\Deps$ error. The difference with respect to the naive expectation $\delta C = -1$ is due to $r^{(8)}_{0,E_0} \neq 0$ (see table {\ref{tab:PBENDR}}).}

\section{\boldmath Beyond Minimal Flavor Violation}\label{sec:beyondMFV}

We would like now to shortly address the case of new physics beyond MFV. Concerning non-MFV contributions to the $\D F = 2$ loop functions, very little can be said 
with present errors on the relevant lattice matrix elements. Indeed, as we have seen in section \ref{sec:loop}, even a universal CMFV shift in the top contribution, 
producing the predictions (\ref{DMdDMsCMFV}), is consistent with experiment as long as lattice matrix element allow for a 20\% uncertainty.

On the other hand, much more can be said on new-physics contributions beyond MFV to $\keps$. 
A first comment concerns the possible impact of the magnetic operators $Q_{11,12}$ \cite{BFG} and new physics therein. These operators affect in principle our strategy in 
two ways. First, they add the unknown parameters $B^{(1/2)}_{11}$, $B^{(1/2)}_{12}$ and $B^{(3/2)}_{12}$. However, since $Q_{11}$ contributes only to the 
$\D I = 1/2$ amplitude, within our strategy its effect is accounted for as a mere shift in the central value of $R_6$. Concerning the $Q_{12}$ matrix elements, they 
are very suppressed, if not vanishing. Hence they can safely be set to zero \cite{BFG}. Second, $Q_{11}$ and $Q_{12}$ mix -- at the two-loop level -- with $Q_{1-10}$. 
In ref. \cite{BFG}, the mixing with the QCD-penguin operators has been estimated as a roughly $10\%$ increase in the $\D I = 1/2$ part of $\epe$. In our case this 
effect can be lumped into the $R_6$ estimate. In other words, similarly to the SM and new-physics effects of the operators $Q_{1-6}$, those of the operators $Q_{11,12}$ 
are taken into account by leaving $R_6$ as a free parameter. Concerning the two-loop QCD mixing between $Q_{11,12}$ and $Q_{7-10}$ \cite{BaranowskiMisiak,GambinoHaisch}, 
as well as the QED one \cite{BGGH,Huber}, to our knowledge no analysis exists exploring their possible impact on $\eps_K$. However, we expect this impact to be well 
within the theoretical error associated with our procedure.

\begin{figure*}[tb]
\begin{center}
\includegraphics[width=0.48 \textwidth]{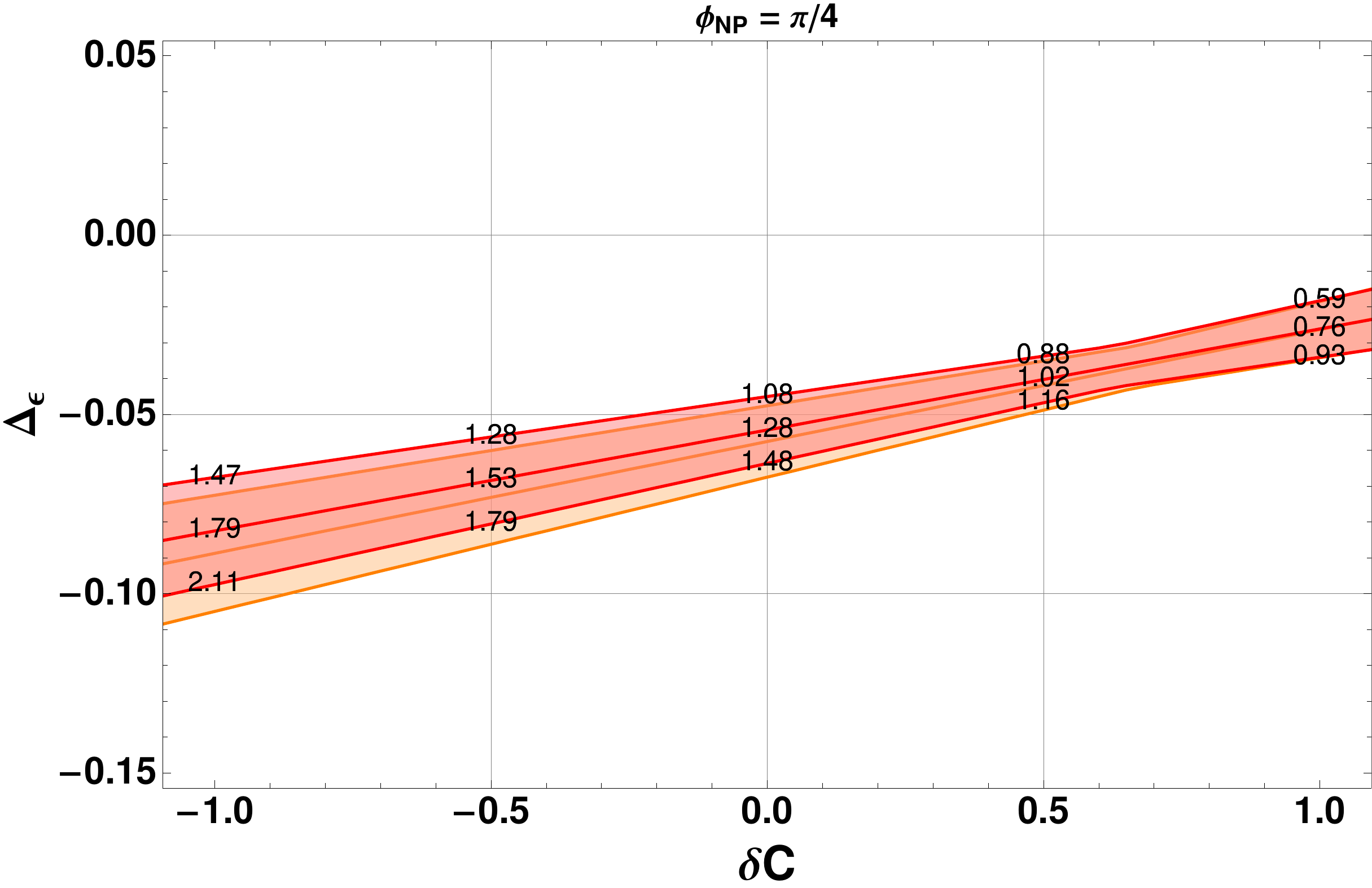} \phantom{xx} \includegraphics[width=0.48 \textwidth]{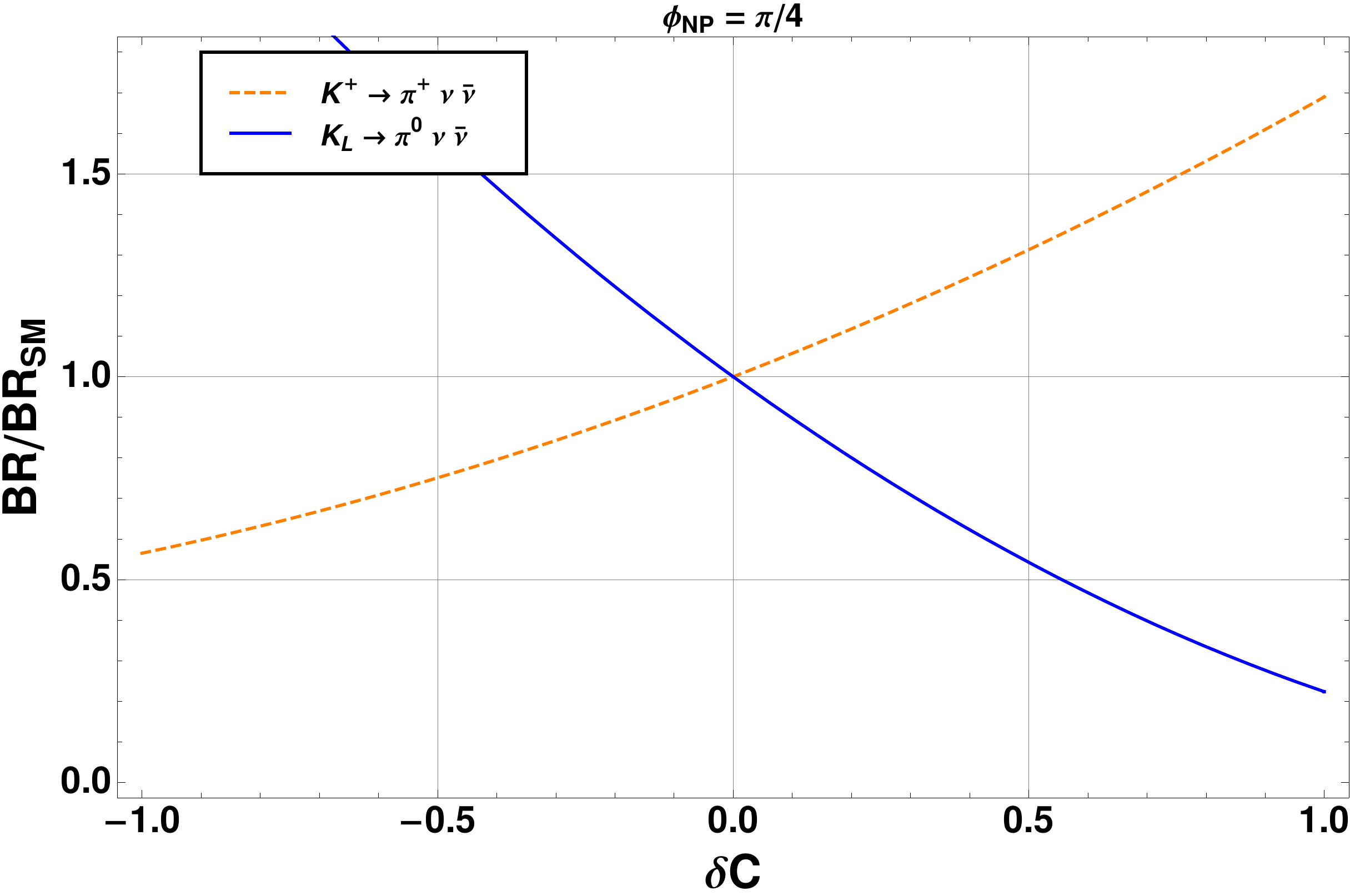}
\includegraphics[width=0.48 \textwidth]{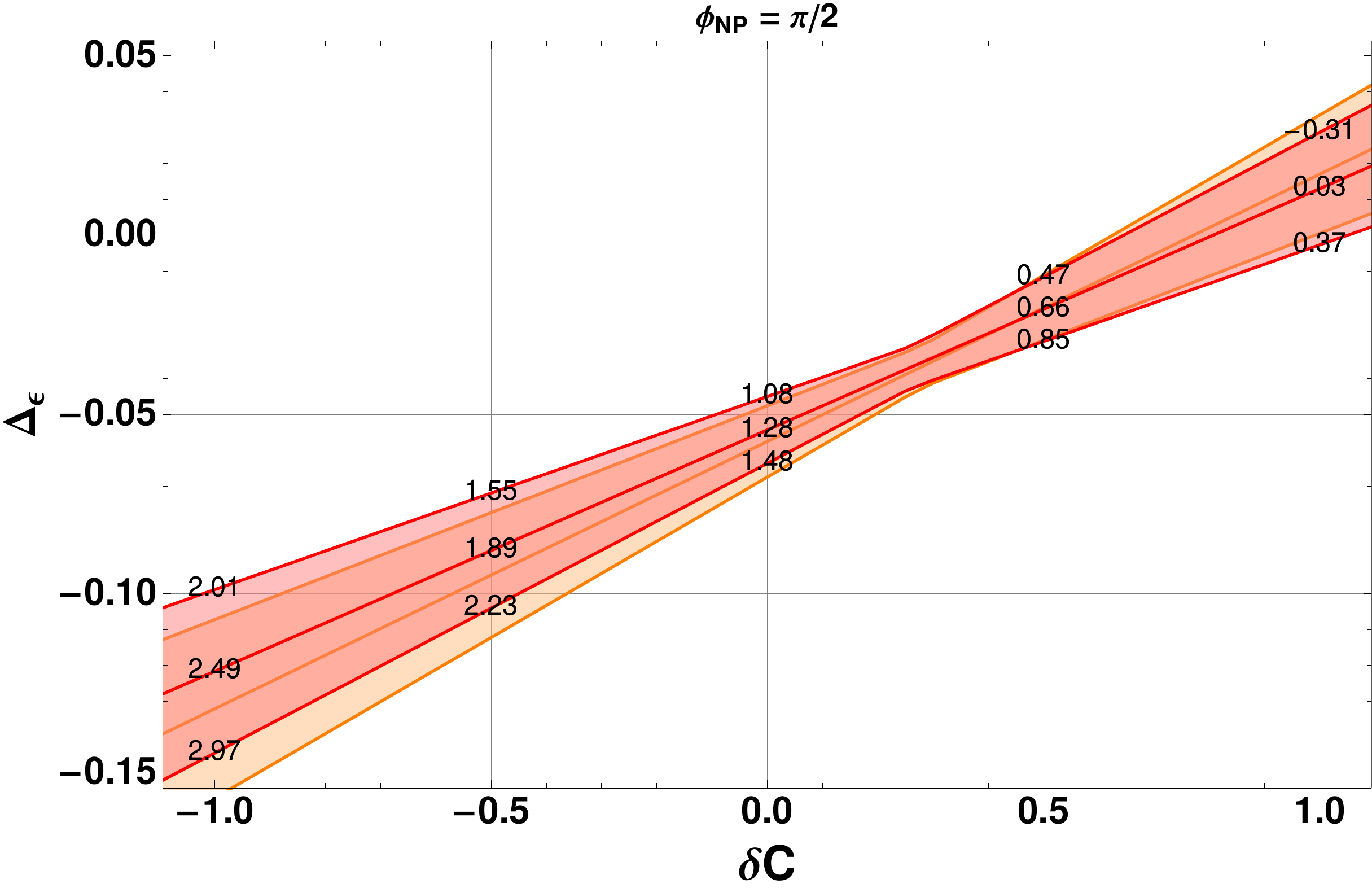} \phantom{xx} \includegraphics[width=0.48 \textwidth]{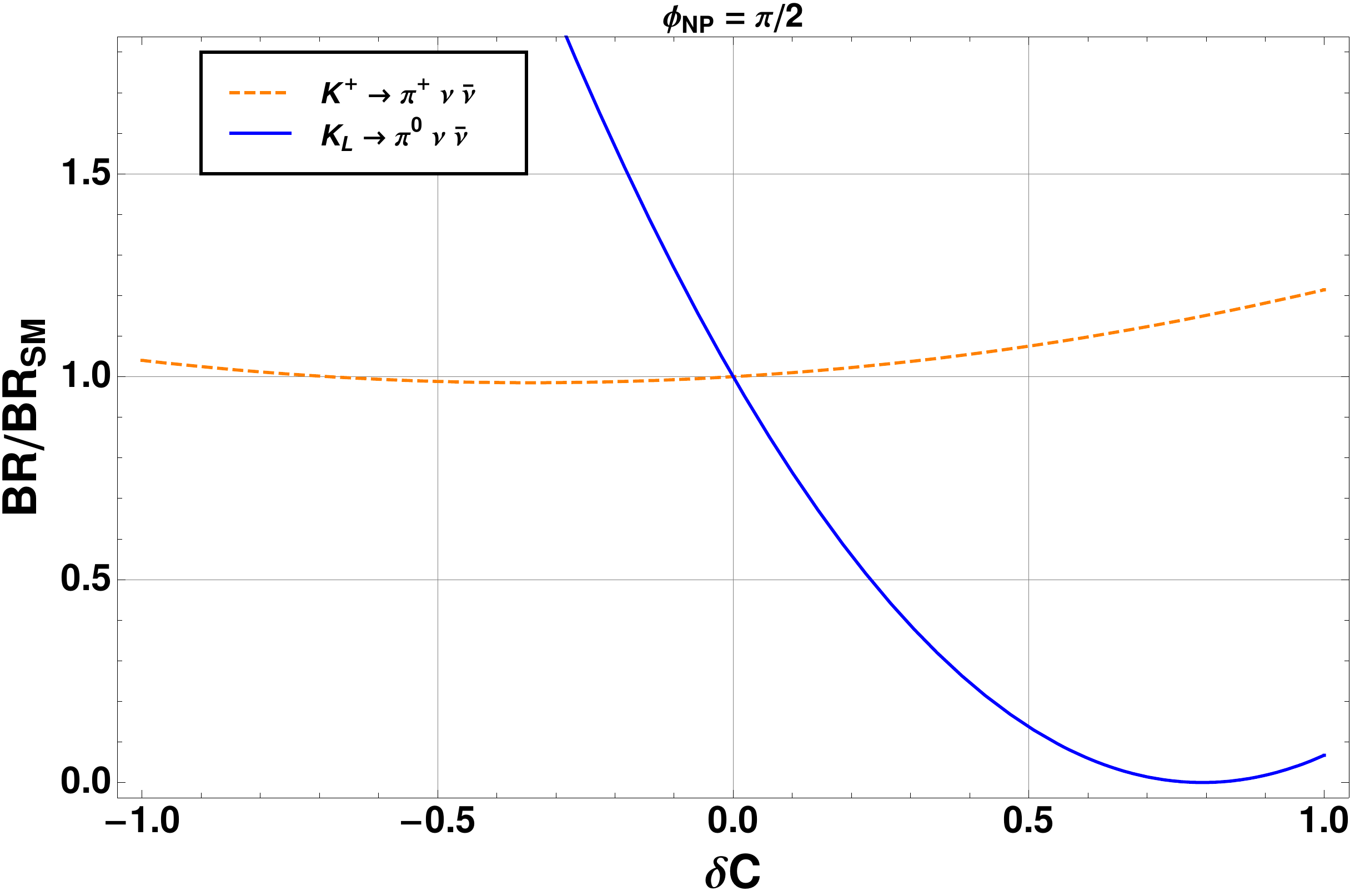}
\includegraphics[width=0.48 \textwidth]{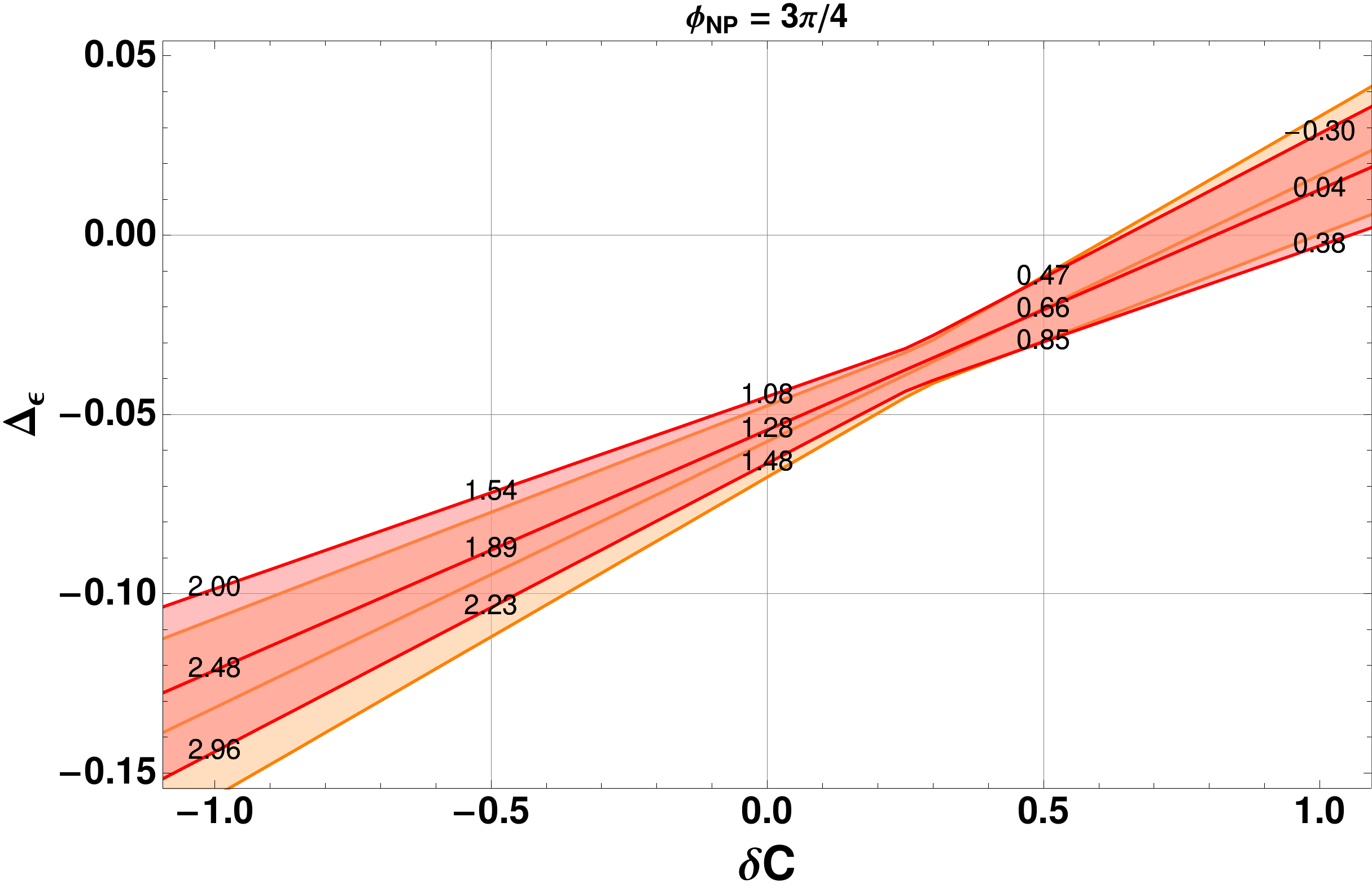} \phantom{xx} \includegraphics[width=0.48 \textwidth]{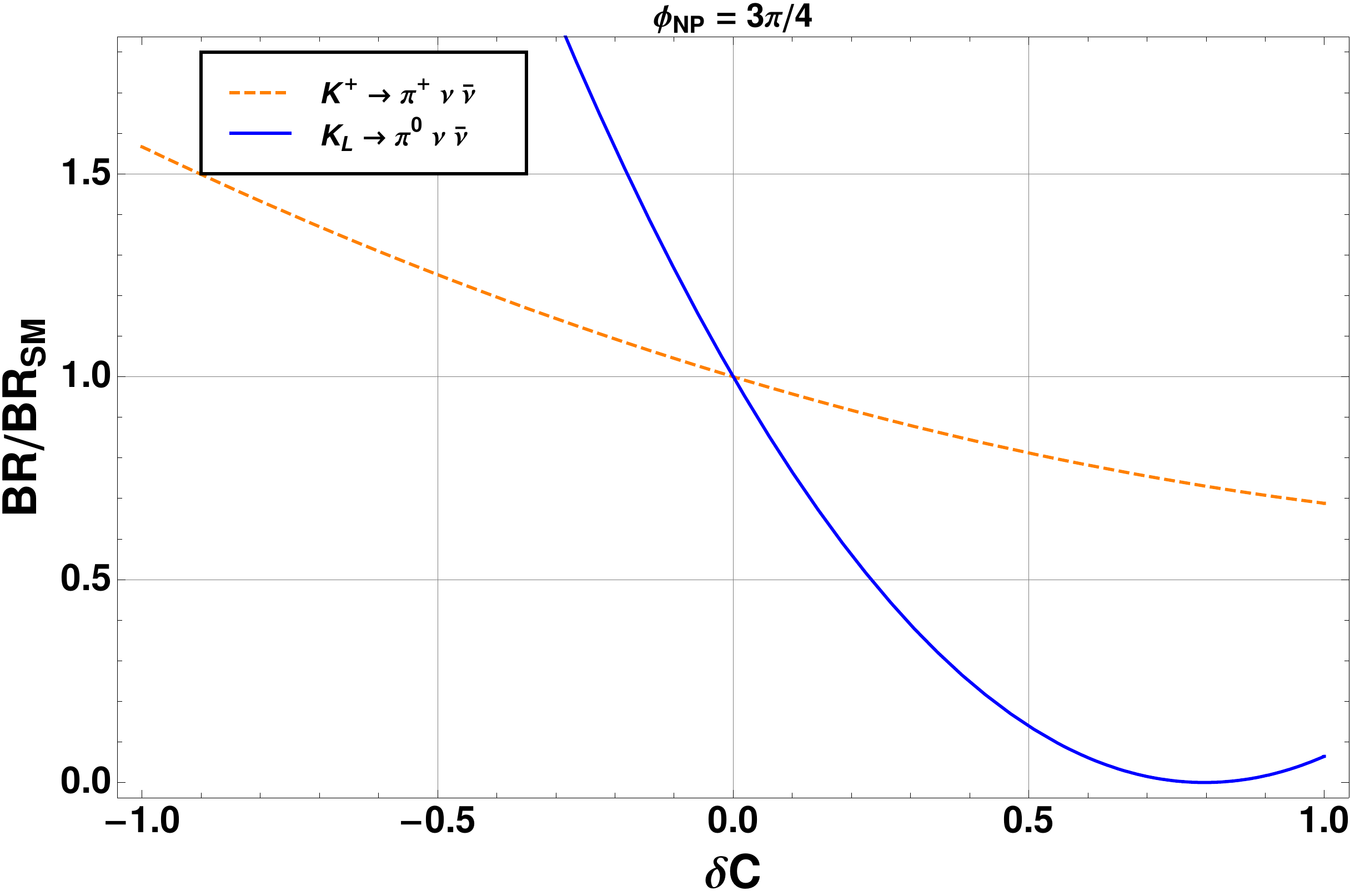}
\end{center}
\caption{\small\sl Same as figure \ref{fig:CvsD}, but for the $Z$-penguin new-physics phase chosen as $\phiNP \in \{\pi/4, \pi/2, 3\pi/4\}$
in the upper, central and lower panels, respectively.}
\label{fig:CvsDnonMFV}
\end{figure*}

A second issue is the possible presence of new phases. With regards to $Z$-penguins, this would amount to $\phiNP \neq 0$ in our parameterization (\ref{ZdsNP}). The ensued
effect on $\Deps$ is displayed in figure \ref{fig:CvsDnonMFV}, which is the analogous of figure \ref{fig:CvsD}, but for the $Z$-penguin new-physics 
phase chosen as $\phiNP \in \{\pi/4, \pi/2, 3\pi/4\}$ in the three figure rows. Plots with $\phiNP$ values in the third and fourth quadrants can obviously be obtained by 
just flipping the $\delta C$ axis.

The right panels of each row in figure \ref{fig:CvsDnonMFV} demonstrate the strong sensitivity of the rate of enhancement for the decays $K^+ \to \pi^+ \nu \ov \nu$ and 
$K_L \to \pi^0 \nu \ov \nu$ to the possible presence of a new phase in $Z$-penguins (cf. \cite{CI} and \cite{BFRS}). The flip side of the coin is however the 
loss of correlation with the $\Deps$ modification, as compared to the $\phiNP = 0$ case of figure \ref{fig:CvsD}. However, a feature that can be read from both 
figs. \ref{fig:CvsD}-\ref{fig:CvsDnonMFV} is that, if one advocates $Z$-penguin contributions to decrease $|\Deps|$, this implies a ratio between 
BR($K^+ \to \pi^+ \nu \ov \nu$) and BR($K_L \to \pi^0 \nu \ov \nu$) larger than in the SM, where this ratio is about 3.

Concerning $\Deps$, one can in addition notice the change in `slope' as a function of the $\delta C$ shift with respect to the left panel of figure \ref{fig:CvsD}. 
This is easy to understand from the following approximate numerical relation for $\epe$:
\beqn
\label{epeNUM}
\frac{1}{\omega}\frac{\epsp}{\eps} &\approx& ... + 0.047 R_6 - 0.018 R_8 \\
&+& (-0.027 \cos \phiNP + 0.067 \sin \phiNP) R_8 \delta C~,\nn
\eeqn
where dots denote other terms, e.g. constant ones, unimportant in this discussion. One can see that, for $\phiNP = 0$, an increase in $\delta C$ implies
an increase in $R_6$ (recall that the r.h.s. of eq. (\ref{epeNUM}) is required to be numerically within the experimental $\epe$ range), i.e. in $|\Deps|$.
However, already for $\phiNP = \pi/4$, the term in the parenthesis on the r.h.s. of eq. (\ref{epeNUM}) has roughly flipped sign, and now an increase in $\delta C$ 
means a decrease in $R_6$, hence in $|\Deps|$.

\section{\boldmath Conclusions}

We have reconsidered the test of compatibility between CP violation in the $K$- and the $B_d$-systems within the SM, by analyzing the $\eps_K$ prediction
implied by $\sin 2 \beta$. As already hinted at by the analysis in \cite{BG}, $\eps_K^{\rm SM}$ can explain only about 80\% of the experimental result, potentially 
signaling an inconsistency, presently masked by a 15\% input uncertainty.

Assuming that the problem be not in the parametric input relevant to $\eps_K$, we have addressed the question whether the mentioned tension could be removed 
without going beyond the MFV framework. The most efficient solution to the tension in question is realised in CMFV, i.e. without advocating operator structures besides 
those relevant in the SM. This solution proceeds through a positive shift in the $\D F = 2$ top-top loop function, and implies $\D M_{d,s}$ predictions roughly $20\%$ 
above experiment. Therefore, with improved determinations of the relevant lattice input, this shift would have to be compensated by decreased values of 
$F^2_{B_q} \hat B_q$ in order for the CMFV predictions to be in agreement with the experimental $\Delta M_{d,s}$. This is illustrated in fig. \ref{fig:FsBvsBK}.

Another avenue would be an increase of the factor $\keps$ in $\eps_K$, that in the SM we estimated to be $\keps = 0.92(2)$. We showed that, within the framework of CMFV,
the needed increase in $\keps$ is correlated, through $\epe$, with a suppression in the branching ratios of $K_L \to \pi^0 \nu \ov \nu$ and $K^+ \to \pi^+ \nu \ov \nu$, 
that is not supported by present -- however limited -- data on the latter decay mode \cite{Kp-exp}. Even admitting this case, we find $\keps \lesssim 0.95$, once other
relevant CMFV constraints \cite{HW} are taken into account, the upper bound holding for a new-physics contribution of O(1) with respect to the SM one. Therefore we conclude
that our SM estimate of $\keps$ is robust also within CMFV at large. Solution to the tension, within the CMFV frameworks, would be a positive shift in the loop function 
$S_0$.

In general MFV frameworks, where new operators matter, addressing the tension between $(\sin 2 \beta)_{J/\psi K_S}$ and $\eps_K$ is a model-dependent issue. However, 
this tension appears to be increased in the case of the MFV MSSM, where contributions from new operators arise for large $\tan \beta$.

Beyond MFV, agreement between $\eps_K$ and $(\sin 2 \beta)_{J/\psi K_S}$ can of course be achieved through appropriate new-physics contributions to the $\D F = 2$ 
Hamiltonians, in general different in the $K$- and $B_d$-systems, and/or through an increase in $\keps$. Figure \ref{fig:CvsDnonMFV} shows the implications on rare $K$ 
decays for a scenario where the $\keps$ increase is due to new physics dominantly in $Z$ penguins.

The possibility to really probe all the above options rests however on improved values of $\bk$, $V_{cb}$ -- on which $\eps_K$ carries strong sensitivity -- as well as of 
$F_{B_q}^2 B_q$, crucial instead for $\D M_q$. The accuracy on these input quantities parametrically rules the accuracy of the consistency test of CP violation between 
the $K$- and the $B_d$-systems within MFV frameworks. A complementary route would be an alternative, direct measurement of the phase in the CKM matrix. That of the UT angle 
$\gamma$ from tree-level decays will be a crucial step forward in this direction.

\acknowledgments

We acknowledge useful comments from Monika Blanke, Uli Haisch, Vittorio Lubicz, Federico Mescia and Luca Silvestrini. This work has been supported in part by the Cluster 
of Excellence ``Origin and Structure of the Universe'' and by the German Bundesministerium f{\"u}r Bildung und Forschung under contract 05HT6WOA. D.G. also acknowledges 
partial support from the A. von Humboldt Stiftung.

\bibliography{keps}

\end{document}

%% file: tabPBENDR.tex
\begin{table*}[t!]
\begin{center}
\begin{tabular}{|c||c|c|c||c|c|c||c|c|c|}
\hline
& \multicolumn{3}{c||}{$\Lms=310~\mev$} &
  \multicolumn{3}{c||}{$\Lms=340~\mev$} &
  \multicolumn{3}{c| }{$\Lms=370~\mev$} \\
& \multicolumn{3}{c||}{$(\alpha_s(M_Z) = 0.117)$} &
  \multicolumn{3}{c||}{$(\alpha_s(M_Z) = 0.119)$} &
  \multicolumn{3}{c| }{$(\alpha_s(M_Z) = 0.121)$} \\
\hline
$i$ & $r_i^{(0)}$ & $r_i^{(6)}$ & $r_i^{(8)}$ &
      $r_i^{(0)}$ & $r_i^{(6)}$ & $r_i^{(8)}$ &
      $r_i^{(0)}$ & $r_i^{(6)}$ & $r_i^{(8)}$ \\
\hline
0 &
   --3.574 &  16.552 &   1.805 &
   --3.602 &  17.887 &   1.677 &
   --3.629 &  19.346 &   1.538 \\
$X_0$ &
     0.574 &   0.030 &       0 &
     0.564 &   0.033 &       0 &
     0.554 &   0.036 &       0 \\
$Y_0$ &
     0.403 &   0.119 &       0 &
     0.392 &   0.127 &       0 &
     0.382 &   0.134 &       0 \\
$Z_0$ &
     0.714 &  --0.023 &  --12.510 &
     0.766 &  --0.024 &  --13.158 &
     0.822 &  --0.026 &  --13.855 \\
$E_0$ &
     0.213 &  --1.909 &   0.550 &
     0.202 &  --2.017 &   0.589 &
     0.190 &  --2.131 &   0.631 \\
\hline
\end{tabular}
\end{center}
\caption[]{The coefficients $r_i^{(0)}$, $r_i^{(6)}$ and $r_i^{(8)}$ of formula (\ref{PiPBE}) for various $\Lms$ in the NDR scheme. Taken from
ref. \cite{Buras-Jamin}.}
\label{tab:PBENDR}
\end{table*}